\documentclass[%
 aip,
 amsmath,
 amssymb,
 reprint,%
]{revtex4-1}

\usepackage{amsfonts}
\usepackage{amssymb}

\usepackage{graphicx}
\usepackage{amsmath}
\usepackage{bm}
\DeclareMathAlphabet{\mathcal}{OMS}{cmsy}{m}{n}

\usepackage[english]{babel}
\usepackage{color}
\usepackage[version=3]{mhchem} 

\newcommand{\ket}[1]{\bigl | #1 \bigr \rangle}
\newcommand{\bra}[1]{\bigl\langle #1 \bigr|}
\newcommand{\braket}[2]{\langle #1 | #2 \rangle}

\usepackage{dcolumn}

\usepackage[utf8]{inputenc}
\usepackage[T1]{fontenc}
\usepackage{mathptmx}
\usepackage{etoolbox}

\makeatletter
\def\@email#1#2{%
 \endgroup
 \patchcmd{\titleblock@produce}
  {\frontmatter@RRAPformat}
  {\frontmatter@RRAPformat{\produce@RRAP{*#1\href{mailto:#2}{#2}}}\frontmatter@RRAPformat}
  {}{}
}%
\makeatother
\begin{document}

\title{Simulation of absorption spectra of molecular aggregates: a Hierarchy of Stochastic Pure States approach}
\author{Lipeng Chen}
\affiliation{Max Planck Institute for the Physics of Complex Systems, N\"{o}thnitzer Str 38, Dresden, Germany}
\email{lchen@pks.mpg.de}
\author{Doran I. G. Bennett}
\affiliation{Department of Chemistry, Southern Methodist University, PO Box 750314, Dallas, TX, USA}
\email{doranb@smu.edu}
\author{Alexander Eisfeld}
\affiliation{Max Planck Institute for the Physics of Complex Systems, N\"{o}thnitzer Str 38, Dresden, Germany}
\email{eisfeld@pks.mpg.de}

\begin{abstract}
The simulation of spectroscopic observables for molecular aggregates with strong and structured coupling of electronic excitation to vibrational degrees of freedom is an important but challenging task. The hierarchy of pure states (HOPS) provides a formally exact solution based on local, stochastic trajectories. Exploiting the localization of HOPS for the simulation of absorption spectra in large aggregares requires a formulation in terms of normalized trajectories. 
Here we provide a normalized dyadic equation where the \textit{ket}- and \textit{bra}-states are propagated in different electronic Hilbert spaces. 
This work opens the door to apply adaptive HOPS methods for the simulation of absorption spectra and also to a formulation for non-linear spectroscopy that is perturbative with respect to interactions with the electric field. 
\end{abstract}

\maketitle

\section{Introduction}
The photophysics of molecular aggregates plays a central role in light harvesting by both artificial materials and photosynthetic organisms. \cite{ReviewPS1,ReviewPS2,ReviewPS3,ReviewPS4} 
Optical absorption is a simple, yet powerful, probe of the delocalized excited states (excitons) formed in molecular aggregates. 
The basic molecular exciton theory treats each chromophore as an electronic two-level (or few-level) system coupled to molecular vibrations and a continuum of vibrational degrees of freedom which are responsible for energy dissipation and electronic dephasing (described by the spectral density).\cite{ReviewPS2,SD1,SD2,SD3,SD4}
When the coupling between the electronic states of the chromophore and the system-environment coupling are of similar magnitudes, then perturbative treatments are inappropriate; when the vibrational modes are highly structured it is essential to account for non-Markovian effects. 
From a theoretical point of view, it is thus of great importance to develop approaches which are capable of treating exciton dynamics at finite temperature in a non-perturbative and non-Markovian manner.

While open quantum system methods appropriate to molecular excitons have typically used density matrices,\cite{NPNMRMP, OQS1, OQS2, Tanimura2006JPSJ, Tanimura2020JCP,Makri1995JCP1,Makri1995JCP2,VarMaster1,VarMaster2} wave function methods have been gaining popularity. 
Beside methods that simultaneously propagate electronic and vibrational degrees of freedom (see e.g Refs.~\cite{MCTDH1, MCTDH2, MCTDH3, GMCTDH, MCG, DA1, DA2, MCE1, MCE2} ),
it is also possible to propagate only the relevant electronic degrees of freedom (or occasionally vibronic) subject to a stochastic force term that is constructed to ensure the ensemble reproduces the correct density matrix.
One of these stochasic approaches is the  non-Markovian quantum state diffusion (NMQSD).\cite{NMQSD1,NMQSD2,NMQSD3}
The evolution equations in NMQSD come in two flavours: a linear equation, where individual wave functions are not normalized, and a non-linear one, where each wave function is properly normalized ensuring more efficient convergence of the ensemble.

The hierarchy of pure states (HOPS) \cite{HOPS1,HOPS2} is a  formally exact way to handle the general NMQSD equation. 
In HOPS the non-Markovian influence of the bath degrees of freedom is captured by a hierarchy of vectors in the system Hilbert space that evolve subject to a Gaussian stochastic noise that shares a correlation function with the environment. 
Averaging over the stochastic noise trajectories reproduces the complete system dynamics.
One appealing feature of NMQSD and HOPS is that dynamics are localized by interaction with the thermal environment which allows adaptive algorithms to calculate dynamics in molecules aggregates composed of thousands of particles.\cite{AdpGao,AdpDoran}

The ability to connect exciton dynamics as obtained by HOPS with spectroscopic observables is essential. 
For HOPS calculations, the linear absorption spectrum can be calculated exactly using a single trajectory by setting the noise trajectory equal to zero. 
The full cancellation of noise terms, however, does not directly extend to non-linear spectroscopic calculations.
Furthermore, the zero-noise trajectory does not show dynamic localization and therefore cannot be solved using efficient adaptive algorithms which exploit the localization of single trajectories. 
These constraints greatly limits the applicability of HOPS for simulating non-linear spectroscopy or even linear spectra of large molecular aggregates. 

\begin{figure}
    \centering
    \includegraphics[width=5cm]{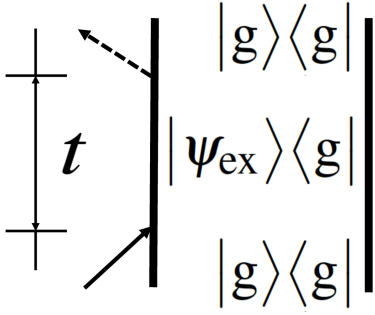}
    \caption{Sketch of the dyadic scheme to calculate linear absorption. 
    Initially the aggregate is in the electronic ground state g, represented by the density matrix $\ket{\mathrm{g}}\bra{\mathrm{g}}$.
    Interactions with the electromagnetic field are indicated by incoming and outgoing arrows.
    The dipole operator then brings the ket into the excited state $\ket{\psi_{\mathrm{ex}}}$; the density matrix is in the state $\ket{\psi_{\mathrm{ex}}}\bra{\mathrm{g}}$.
    Finally, the dipole operator brings the excited state back to the electronic ground state, and the trace is taken.
    For further details about the formalism see Ref.~\onlinecite{MukamelTextBook}
    }
    \label{fig:Dyadic-Sketch}
\end{figure}

Here, we provide a framework for simulating linear absorption using {\it stochastic} HOPS equations. 
One particular goal is to connect the HOPS approach, where pure states are propagated to the common formulation of spectroscopy in terms of dyadic states, where the {\it bra} and {\it ket} can be in different electronic states and are propagated separately.
Such a dyadic propagation scheme is symbolically displayed in the form of a so-called double-sided Feynman diagram, as displayed in Fig.~\ref{fig:Dyadic-Sketch}.
We demonstrate that by expressing the density matrix after the first interaction with an electric field as a sum over pure states, we can write a simple equation that is equivalent to propagating the {\it bra} and {\it ket} side of the initial density matrix separately according to HOPS. 
For a dyadic propagation scheme it is not obvious how to construct the normalization of the non-linear NMQSD equation.
We find an correct normalization contains contributions from both the {\it bra} and {\it ket}, coupling the two time-evolution equations.
We study in detail the convergence of the dipole-dipole correlation function with the number of stochastic trajectories using both the linear and non-linear HOPS equations. 
We find that both, the linear and the non-linear equation are well behaved at the single trajectory level and show good convergence properties. 
Remarkably, this hold even in regimes where the linear HOPS  fails catastrophically for the calculation of site populations.

The paper is organized as follows:
In section \ref{sec:background} we introduce the open quantum system formulation of the molecular aggregate and provide the NMQSD framework and the linear and the non-linear HOPS equations.
In section \ref{sec:spectra}, which contains our central results, we formulate absorption in terms of stochastic NMQSD trajectories and we establish a dyadic HOPS formulation.
In section \ref{sec:calculations} we demonstrate the applicability of our formulas and investigate the convergence with respect to the number of trajectories.
Finally, in section \ref{sec:conclusions} we summarize our findings and we conclude with an outlook.

Throughout the work we use $\hbar=1$.
    
\section{Background\label{sec:background}}
\subsection{Open quantum system description of the aggregate\label{sec:open_quantum}}
We consider a molecular aggregate consisting of $N$ interacting chromophores embedded in a condensed phase environment such as a solvent, protein, or solid matrix. In the language of open quantum system, the Hamiltonian for the aggregate can be written as \cite{ReviewPS2}
\begin{equation}\label{eq:Htot}
\hat{H}=\hat{H}_{\mathrm{S}}+\hat{H}_{\mathrm{B}}+\hat{H}_{\mathrm{SB}}
\end{equation}
with $\hat{H}_{\mathrm{S}}$, $\hat{H}_{\mathrm{B}}$, and  $\hat{H}_{\mathrm{SB}}$ denoting the system Hamiltonian, the environment (bath) Hamiltonian, and the system-bath interaction Hamiltonian, respectively. The system Hamiltonian $\hat{H}_{\mathrm{S}}$ describing the Frenkel-exciton reads 
\begin{equation}
\label{eq:H_sys}
\hat{H}_{\mathrm{S}}=\hat{H}_\mathrm{g}+ \hat{H}_\mathrm{ex}
\end{equation}
with the ground state and excited state Hamiltonians
\begin{eqnarray}
\hat{H}_{\mathrm{g}}&=&\epsilon_\mathrm{g}|\mathrm{g}\rangle\langle{\mathrm{g}}|\\
\nonumber\\ 
\hat{H}_{\mathrm{ex}}&=&\sum_{n=1}^{\mathrm{N}}\epsilon_n|n\rangle\langle{n}|+\sum_{n\neq{m}}^{\mathrm{N}}V_{nm}|n\rangle\langle{m}|,
\end{eqnarray}
where $\epsilon_\mathrm{g}$  is the energy in the ground state, $\epsilon_n$ is the site energy of the $n$th molecule, $|n\rangle$ is the state describing the $n$th excited molecule, and $V_{nm}$ is the electronic coupling between excited states of molecules $n$ and $m$. 
Each chromophore is coupled to inter- and intra-molecular vibrations which can be modeled as a collection of harmonic oscillators. 
In what follows, we assume each molecule is coupled to an independent collection of vibrations.
We thus have the bath Hamiltonian 
\begin{equation}
\hat{H}_{\mathrm{B}}=\sum_{n=1}^N\sum_q  \omega_{nq}\hat{b}_{nq}^{\dagger}\hat{b}_{nq}
\end{equation}
where $\hat{b}_{nq}^{\dagger} \textrm{ } (\hat{b}_{nq})$ are the creation (annihilation) operator of $q$th mode of chromophore $n$ with frequency $\omega_{nq}$. 
The system-bath coupling Hamiltonian $\hat{H}_{\mathrm{SB}}$ is expressed as 
\begin{equation}
\hat{H}_{\mathrm{SB}}=-\sum_{n=1}^N \hat{L}_n \sum_qg_{nq}\left(\hat{b}_{nq}^{\dagger}+\hat{b}_{nq}\right)
\end{equation}
with system coupling operators
\begin{equation}
\hat{L}_n= \ket{n}\bra{n}
\end{equation}
and $g_{nq}$ is the exciton-bath coupling strength of the $q$th bath mode for chromophore $n$. The influence of the vibrational modes on the dynamics of the electronic system is described by the bath-correlation function
\begin{equation}
\alpha_n(\tau)=\int_0^{\infty}\!\!\!\mathrm{d}\omega\, J_n(\omega)\Big( \coth\big(\frac{\beta \omega}{2}\big) \cos (\omega \tau) -i \sin(\omega \tau)\Big)
\end{equation}
that contains the spectral density $J_n(\omega) = \sum_{q}|g_{nq}|^2\delta(\omega-\omega_{nq})$ and the inverse temperature $\beta=1/T$.

\subsection{Non-Markovian quantum state diffusion (NMQSD) and the Hierarchy of Pure States (HOPS)}\label{PropSchemes}
For a separable initial state 
\begin{equation}
\label{eq:rho_ini}
\hat{\rho}_\mathrm{ini}= \ket{\psi_\mathrm{ini}}\bra{\psi_\mathrm{ini}} \otimes \hat{\rho}_\mathrm{B}
\end{equation} 
the expectation value of a system operator $\hat{F}$ can be expressed as
\begin{equation}
\label{eq:expectationVal}
\langle \hat{F}(t) \rangle = {\mathrm{Tr}}[\hat{F} \hat{\rho}(t)]=\mathcal{M}[\bra{\psi(t,\mathbf{z}^*)}\hat{F} \ket{\psi(t,\mathbf{z}^*)}]
\end{equation}
where $\mathcal{M}[\cdots]$ denotes ensemble average over realizations of stochastic trajectories which obey the (linear) non-Markovian quantum state diffusion (NMQSD) equation \cite{NMQSD1,NMQSD2}
\begin{equation}\label{eq:dotPsit_NMQSD}
\begin{split}
\partial_t|\psi(t,\mathbf{z}^{*})\rangle=&-i\hat{H}_{\mathrm{S}}|\psi(t,\mathbf{z}^{*})\rangle\\
&+\sum_n \hat{L}_nz_{t,n}^{*}|\psi(t,\mathbf{z}^{*})\rangle\\
&-\sum_n \hat{L}_n^{\dagger}\int_0^t\mathrm{d}s\,\alpha_n(t-s)\frac{\delta{|}\psi(t,\mathbf{z}^{*})\rangle}{\delta{z}_{s,n}^{*}}
\end{split}
\end{equation} 
where $|\psi(t,\mathbf{z}^*)\rangle$ is a vector in the system Hilbert space and $\mathbf{z}$ comprises complex Gaussian stochastic processes $z_{t,n}^{*}$ with mean $\mathcal{M}[z_{t,n}]=0$, and correlations $\mathcal{M}[z_{t,n}z_{s,m}]=0$, and $\mathcal{M}[z_{t,n}z_{s,m}^{*}]=\alpha_n(t-s)\delta_{nm}$.
The important non-linear NMQSD equation is obtained by making in the above equation the following replacements:\cite{NMQSD2}  $\hat{L}_n^{\dagger}\rightarrow \hat{L}_n^{\dagger}-\langle\hat{L}_n^{\dagger} \rangle_t$  and $\tilde{z}_{t,n}=z_{t,n}^{*}+\int_0^t\mathrm{d}s\,\alpha_n^{*}(t-s)\langle{\hat{L}}_n^{\dagger}\rangle_s$. 
Expectation values  $\langle\cdot\rangle_t$ are calculated using the normalized state.

The hierarchy of pure states (HOPS) is a formally exact solution to the open quantum system model discussed in the previous sections within the NMQSD framework.  
HOPS requires the bath correlation function $\alpha_n(\tau)$ to be expanded as a sum of exponentials 
\begin{equation}\label{BCF}
\alpha_n(\tau) \approx \sum_{j=1}^Jp_{nj}e^{-w_{nj}\tau}
\end{equation} 
where $w_{nj}=\gamma_{nj}+i\Omega_{nj}$. 
Assuming, each molecule has the same bath correlation function, then the linear HOPS equation is
\begin{eqnarray}
\label{LinHOPSEq}
\begin{aligned}
\partial_t|\psi^{(\mathbf{k})}&(t,\mathbf{z}^{*})\rangle
\\
=
&\Big(-i\hat{H}_{\mathrm{S}}-\mathbf{k}\cdot\mathbf{w}+\sum_n\hat{L}_nz_{t,n}^{*}\Big)|\psi^{(\mathbf{k})}(t,\mathbf{z}^{*})\rangle\\
&+\sum_{n}\hat{L}_n \sum_j k_{nj}p_{nj}|\psi^{(\mathbf{k}-\mathbf{e}_{nj})}(t,\mathbf{z}^{*})\rangle \\
&-\sum_{n}\hat{L}_n^{\dagger}\sum_j|\psi^{(\mathbf{k}+\mathbf{e}_{nj})}(t,\mathbf{z}^{*})\rangle 
\end{aligned}
\end{eqnarray}
where $\mathbf{w}=\left\lbrace{w}_{1,1},\cdots,w_{N,J}\right\rbrace$,
and $\mathbf{k}=\left\lbrace{k}_{1,1},\cdots,k_{N,J}\right\rbrace$ with non-negative integers $k_{nj}$.
The appearance of $\mathbf{z}$ in  $|\psi(t,\mathbf{z}^*)\rangle$ indicates that the state depends on the complete set of stochastic processes up to time $t$. 
With ${}^*$ we denote complex conjugate.
The physical wave function is given by $|\psi^{(\mathbf{0})}(t,\mathbf{z}^{*})\rangle$ while the remaining terms are `auxiliary wave functions' and represent the influence of the finite memory time of the bath on the time evolution of the electronic state. 
The time evolution of the reduced density matrix associated with the system degrees of freedom is given by $\hat{\rho}(t) = \mathcal{M}\big[|\psi^{(\mathbf{0})}(t,\mathbf{z}^{*})\rangle \langle \psi^{(\mathbf{0})}(t,\mathbf{z}^{*})|\big]$.
HOPS consists of infinite set of coupled equations, which can be truncated at a finite number of hierarchy elements, for example using the triangular truncation condition for the hierarchy where only auxiliary states with $|\mathbf{k}|\leq\mathcal{K}$ are taken into account, more flexible scheme to truncate the HOPS can be found in Ref.~\onlinecite{TrunHOPS}. 

The non-linear version of the HOPS equation reads
\begin{eqnarray}\label{NonLinHOPSEq}
\begin{aligned}
\partial_t|\tilde{\psi}^{(\mathbf{k})}(t,\mathbf{z}^{*})\rangle=&\Big(-i\hat{H}_{\mathrm{S}}-\mathbf{k}\cdot\mathbf{w}+\sum_n \hat{L}_n\tilde{z}_{t,n}\Big)|\tilde{\psi}^{(\mathbf{k})}(t,\mathbf{z}^{*})\rangle\\
&+\sum_{n}\hat{L}_n \sum_j k_{nj}p_{nj}|\tilde{\psi}^{(\mathbf{k}-\mathbf{e}_{nj})}(t,\mathbf{z}^{*})\rangle  \\
&-\sum_{n}\left(\hat{L}_n^{\dagger}-\langle{\hat{L}}_n^{\dagger}\rangle_t\right)\sum_j |\tilde{\psi}^{(\mathbf{k}+\mathbf{e}_{nj})}(t,\mathbf{z}^{*})\rangle.
\end{aligned}
\end{eqnarray}
Here, $\tilde{z}_{t,n}=z_{t,n}^{*}+\int_0^t\mathrm{d}s\,\alpha_n^{*}(t-s)\langle{\hat{L}}_n^{\dagger}\rangle_s$. 
Expectation values  $\langle\cdot\rangle_t$ are calculated using the normalized physical state 
\begin{equation}
\label{eq:Psi_normalized}
    \ket{\Psi^{(\mathbf{0})}(t,\mathbf{z}^*)}
    =\frac{\ket{\tilde{\psi}^{(\mathbf{0})}(t,\mathbf{z}^{*})}}
    {
    \sqrt{\braket{\tilde{\psi}^{(\mathbf{0})}(t,\mathbf{z}^{*})}
    {\tilde{\psi}^{(\mathbf{0})}(t,\mathbf{z}^{*})}
    }}.
\end{equation}
We emphasize that in particular for strong coupling the non-linear equation is essential to ensure convergence. 
Also for adaptivity the non-linear normalized equation is very beneficial.

\section{Linear spectra from stochastic trajectories\label{sec:spectra}}
To calculate absorption spectra of the aggregate, we need to extend the Hamiltonian to include the interaction of the electronic system with the external electric field \cite{MukamelTextBook}
\begin{eqnarray}
\hat{H}_{\mathrm{L}}
&=&-\sum_n^N\big((\boldsymbol{\mu}_n\cdot\boldsymbol{\varepsilon})E(t)\big)|n\rangle\langle{\mathrm{g}}|+h.c.
\end{eqnarray} 
where $\boldsymbol{\mu}_n$ is the transition dipole moment of chromophore $n$, and $\boldsymbol{\varepsilon}$ and $E(t)$ are the polarization and the envelope of the laser pulse, which we have taken to be constant over the size of the aggregate. Our results can be easily extended to electric field that vary from molecule to molecule, as needed for example for circular dichroism \cite{FTODTFD} or for near-field spectroscopy.\cite{NFS1,NFS2}

The linear absorption spectra ($\mathcal{A}(\omega)$) is given by the Fourier transformation 
\begin{equation}
\mathcal{A}(\omega)=\mathrm{Re}\int_0^{\infty}\mathrm{d}t \, e^{i\omega{t}}C(t)
\end{equation}
of the dipole-dipole auto-correlation function \cite{MukamelTextBook}
\begin{eqnarray}\label{eq:Ct_def}
C(t)
=\mathrm{Tr}\left\lbrace\hat{\mu}_{\mathrm{eff}}\,{e}^{-i\hat{H}t}\big(\hat{\mu}_\mathrm{eff}{|}\mathrm{g}\rangle\langle{\mathrm{g}}|\otimes\hat{\rho}_{\mathrm{B}}\big) e^{i\hat{H}t}\right\rbrace
\end{eqnarray}
where $\hat{\mu}_\mathrm{eff}$ is the scalar, collective dipole moment operator 
\begin{equation}
\label{eq:collective_Transition_Operator}
\hat{\mu}_\mathrm{eff}=\sum_n^N (\boldsymbol{\mu}_n\cdot\boldsymbol{\varepsilon}) \,|n\rangle\langle{\mathrm{g}}|+ h.c.
\end{equation}
and $\hat{\rho}_0=|\mathrm{g}\rangle\langle{\mathrm{g}}|\otimes\hat{\rho}_{\mathrm{B}}$ is a factorized initial total density matrix with $\hat{\rho}_{\mathrm{B}}=e^{-\beta{\hat{H}}_{\mathrm{B}}}/\mathrm{Tr}_{\mathrm{B}}\left\lbrace{e}^{-\beta{\hat{H}}_{\mathrm{B}}}\right\rbrace$ being the density matrix of the thermal bath.

\subsection{Pure state decomposition of initial density matrix}
The NMQSD framework cannot directly applied to the expression (Eq. \ref{eq:Ct_def}), since the initial state  $\hat\mu_\mathrm{eff} \ket{\mathrm{g}} \bra{\mathrm{g}}$ is not pure.
To construct the correlation function in terms of pure states, we follow the approach of Hartmann and Strunz.\cite{HOPSEDRichard} 
We find the decomposition
\begin{equation}
\label{eq:pure_decomp}
\hat{\mu}_\mathrm{eff} \ket{\mathrm{g}}\bra{\mathrm{g}}=
\frac{1}{2}\, \mu_\mathrm{tot} \sum_{\eta\in \{\pm 1, \pm i\}} \eta \ \ket{v_{\eta}}\bra{v_{\eta}}
\end{equation}
into pure states given by
\begin{equation}
\label{eq:initial_states_diag}
\ket{v_{\eta}}=\frac{1}{\sqrt{2}} (\ket{\psi_\mathrm{ex}}+ \eta\ket{\mathrm{g}}), \quad\quad \eta\in \{\pm 1, \pm i\}
\end{equation}
where
\begin{equation}
\label{eq:psi_ex}
\ket{\psi_\mathrm{ex}}
= 
\frac{1}{{\mu}_\mathrm{tot}} \hat{\mu}_\mathrm{eff} \ket{\mathrm{g}} 
=
\frac{1}{{\mu}_\mathrm{tot}} \sum_{n=1}^N (\boldsymbol{\mu}_n \cdot \boldsymbol{\varepsilon}) \ket{n}
\end{equation}
\begin{equation}
\mu_\mathrm{tot}= \sqrt{\sum_{n=1}^N (\boldsymbol{\mu}_n \cdot \boldsymbol{\varepsilon})^2 }
\end{equation}
and $ \mu_\mathrm{tot}$ is the total transition strength. 

With this pure state decomposition, the dipole-dipole auto-correlation function can be written as a sum over contributions from each pure state
and accordingly each pure state contribution can be treated within the NMQSD framework. 
Using Eq.~(\ref{eq:expectationVal}) one finds
\begin{equation}
\label{eq:C(t)_diag}
C(t)= \frac{1}{2} \mu_\mathrm{tot} \sum_{\eta\in \{\pm 1, \pm i\}} \eta\,\mathcal{M} [\bra{v_\eta(t,z^*)} \hat\mu_\mathrm{eff} \ket{v_\eta(t,z^*)}  ]
\end{equation}
where $\ket{v_\eta(t,z^*)}$ denotes the time evolved states $\ket{v_\eta}$ according to the NMQSD equation in the full Hilbert space.

\subsection{Direct propagation of the dyadic equation}
The correlation function can be calculated by directly propagating a single initial state defined in the one exciton manifold. We will demonstrate that this method is equivalent to propagating a dyadic equation where the bra and ket states are independently propagated to construct the correlation function. 

First, we will simplify the time evolution of the pure state components of the correlation function to separate the excited state and ground state dynamics. We note that the Hamiltonian (Eq. \ref{eq:Htot}) does not couple the electronic ground state $\ket{\mathrm{g}}$ and the single exciton states $\ket{n}$: the system Hamiltonian has the form $\hat{H}_\mathrm{S}=\hat{H}_\mathrm{g} + \hat{H}_\mathrm{ex}$ and the system-bath coupling Hamiltonian acts only in the one-exciton space. Therefore, initial states of the form of Eq.~\ref{eq:initial_states_diag} will evolve under the NMQSD equation to give
\begin{equation}
\label{eq:v_eta(t)}
\ket{v_\eta(t,z^*)}= \frac{1}{\sqrt{2}}\big(\ket{\psi_\mathrm{ex}(t,z^*)}+ \eta e^{-i \epsilon_{\mathrm{g}} t} \ket{\mathrm{g}})
\end{equation}
where $\ket{\psi_\mathrm{ex}(t,z^*)}$ denotes the state that is obtained upon evolving the initial state $\ket{\psi_\mathrm{ex}}$ with the NMQSD equation restricted to the one-exciton space (i.e.\ replacing $\hat{H}_\mathrm{S}$ by $\hat{H}_\mathrm{ex}$ in Eq.~(\ref{eq:dotPsit_NMQSD}), and correspondingly in the HOPS equations).

Using this result together with the explicit form of the collective dipole operator (Eq.~(\ref{eq:collective_Transition_Operator})) we find that the terms appearing in Eq.~(\ref{eq:C(t)_diag}) have the form
\begin{equation}
\bra{v_\eta(t,z^*)} \hat\mu_{\mathrm{eff}} \ket{v_\eta(t,z^*)}
=
\frac{1}{2}\eta^* \mu_\mathrm{tot}\, \braket{\psi_\mathrm{ex}}{\psi_\mathrm{ex}(t,z^*)}e^{i \epsilon_{\mathrm{g}} t}  +h.c.
\end{equation}
Recognizing that 
\begin{equation}
\begin{split}
    \eta\,\bra{v_\eta(t,z^*)} &\hat\mu_{\mathrm{eff}} \ket{v_\eta(t,z^*)} 
    \\
=&
\phantom{+}\frac{1}{2} \mu_\mathrm{tot}\,\vert \eta \vert^2 \braket{\psi_\mathrm{ex}}{\psi_\mathrm{ex}(t,z^*)}e^{i \epsilon_{\mathrm{g}} t} 
\\&+\frac{1}{2} \mu_\mathrm{tot}\, \eta^2 \braket{\psi_\mathrm{ex}(t,z^*)}{\psi_\mathrm{ex}}e^{-i \epsilon_{\mathrm{g}} t}
\end{split}
\end{equation}
where $\vert \eta \vert^2 = 1$ for all $\eta = \pm 1, \pm i$ then the four contributions to the correlation function simplify (due to cancellation) to give
\begin{equation}
\label{eq:C(t)_final}
C(t)=  \mu_\mathrm{tot}^2 \mathcal{M}[\braket{\psi_\mathrm{ex}}{\psi_\mathrm{ex}(t,z^*)}]e^{i \epsilon_{\mathrm{g}} t}.
\end{equation}

Eq. \ref{eq:C(t)_final} is our  central result and shows that the dipole correlation function can be calculated by propagating the initial state $\ket{\psi_\mathrm{ex}}=\sum_{n=1}^N \frac{(\boldsymbol{\mu}_n \cdot \boldsymbol{\varepsilon} )}{\mu_{\mathrm{tot}}} \ket{n}$ with the NMQSD equation up to time $t$ and then project on the same initial state again. The average over many such trajectories will reproduce the ensemble correlation function. Note that the form of Eq.~(\ref{eq:C(t)_final}) is very similar to that of the noiseless equation.\cite{FTODTFD} 

It is easy to see (details can be found in appendix \ref{sec:calc_dyadic}) that Eq.~\ref{eq:C(t)_final} can be written as 
\begin{equation}
\label{eq:C(t)_dyadic}
C(t)={\mu}_\mathrm{tot} \, \mathcal{M} \Big[ \mathrm{Tr}_\mathrm{S}\Big\{\hat{\mu}_\mathrm{eff} \ket{\psi_\mathrm{ex}(t,z^*)}\bra{\mathrm{g}(t)}\Big\}\Big]
\end{equation}
with $\ket{\mathrm{g}(t)}=e^{-i \hat{H}_\mathrm{g} t} \ket{\mathrm{g}}$,
which can also be written as
\begin{equation}
C(t)= \mathcal{M} \Big[ \mathrm{Tr}_\mathrm{S}\{\hat{\mu}_\mathrm{eff}\, (\hat{U}_\mathrm{ex}(t,z^*) \hat\mu_\mathrm{eff} \ket{\mathrm{g}}) \bra{\mathrm{g}}\hat{U}^\dagger_\mathrm{g}(t)\} \Big]
\end{equation}
where $\hat{U}_\mathrm{ex}(t,z^*)$ denotes the stochastic propagator in the excited state manifold, i.e.\ evolution with the respective NMQSD equation \cite{NMQSDfootnote}
We see that this equation has now exactly the form of a dyadic scheme as sketched in Fig.~\ref{fig:Dyadic-Sketch}: The \textit{bra} is during the whole time evolution propagated in the electronic ground state. 
The \textit{ket} is first lifted via $\hat\mu_\mathrm{eff}$ to the excited state manifold where it is then propagated during the time period $t$. 
Finally, $\hat\mu_\mathrm{eff}$ is applied again and the trace is taken.
One has to be careful about the formal meaning of the stochastic state $\ket{\psi_\mathrm{ex}(t,z^*)}$ in the case of the non-linear NMQSD equation.
We will discuss this in detail in the next section.

\begin{figure*}[]
\centering
\includegraphics[width=16cm]{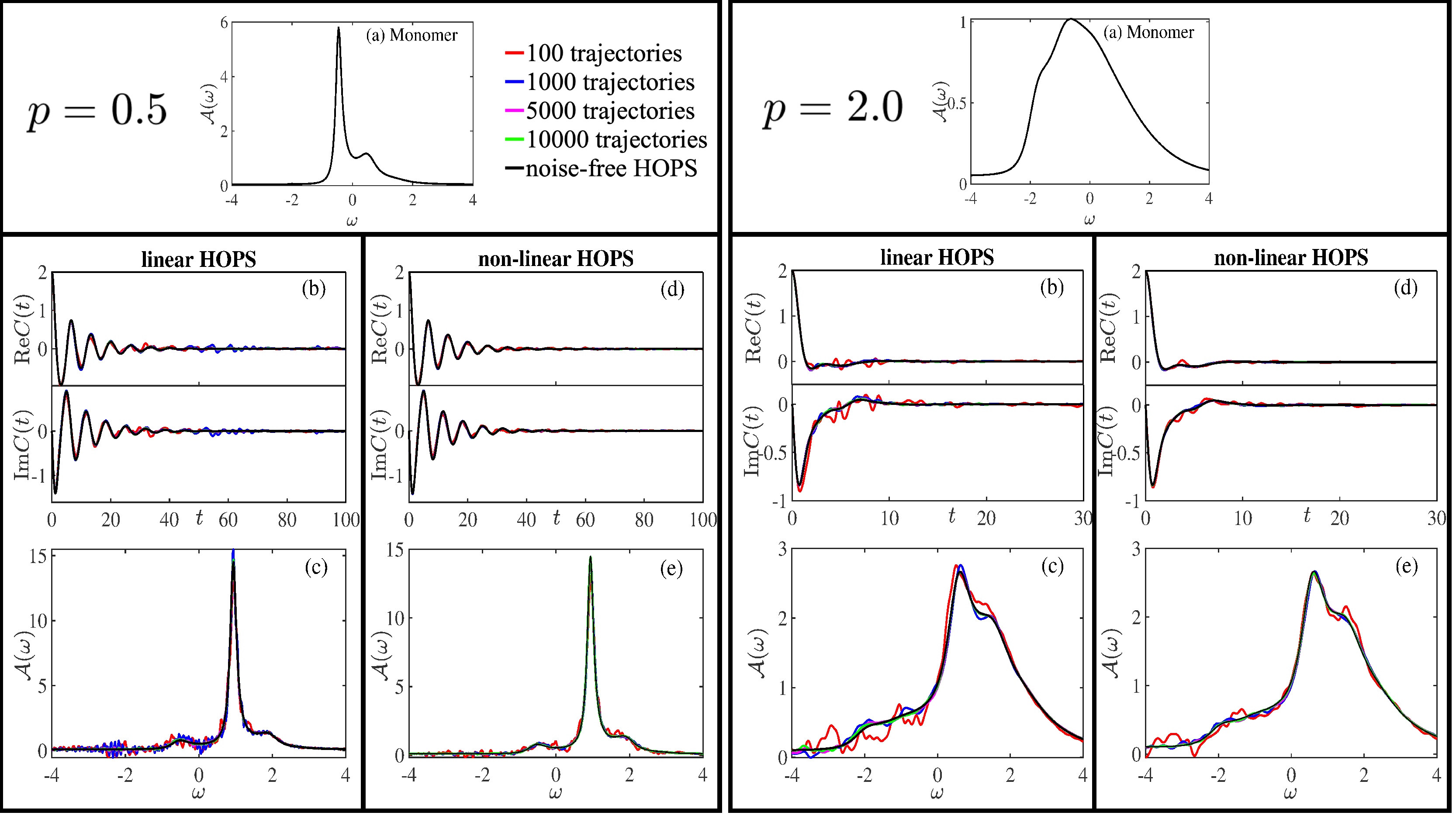}
\caption{
Example calculations for the case of a bath-correlation function described by one exponential with $\Omega=1.0$ and $\gamma=0.25\Omega$. 
Two cases of the coupling strength are shown: intermediate coupling $p=0.5$ (left box) and strong coupling $=2$ (right box).
The dipole-dipole coupling between the monomer is $V=\Omega$. 
In each box we show the following:
(a) Monomer spectrum. (b) real and imaginary parts of $C(t)$, (c) absorption spectrum calculated by the linear HOPS equation. (d) real and imaginary parts of $C(t)$, (e) absorption spectrum calculated by the non-linear HOPS equation.
The truncation depth of the hierarchy for HOPS are $\mathcal{K}=6$ ($p=0.5$) and $\mathcal{K}=12$ ($p=2$). 
The noise-free HOPS calculations are performed using the formulas provided in appendix~\ref{sec:NoisefreeHOPS}. 
For the stochastic calculations Eq.~(\ref{eq:C(t)_final}) and Eq.~(\ref{eq:C(t)_final_normalized}) are used for the linear and non-linear HOPS, respectively.
} 
\label{fig:Example_spectra_1}
\end{figure*}

\subsection{Dyadic non-linear HOPS equation}
While we have demonstrated that the NMQSD equations can be used to directly propagate a dyadic equation for the dipole-dipole auto-correlation function, but care is required to establish the corresponding non-linear HOPS equation. In the non-linear form of the starting equation Eq.~(\ref{eq:C(t)_final}) one propagates the pure state vectors  $\ket{\tilde{v}_\eta(t,z^*)}$ according to Eq.~(\ref{NonLinHOPSEq}) in the total Hilbert space spanned by the singly excited states $\ket{n}$ and the ground state $\ket{\mathrm{g}}$.   
It has the same form as the Eq.~(\ref{eq:v_eta(t)}) $\ket{\tilde{v}_\eta(t,z^*)}= \frac{1}{\sqrt{2}}\big(\ket{\tilde{\psi}_\mathrm{ex}(t,z^*)}+ \eta e^{-i \epsilon_{\mathrm{g}} t} \ket{\mathrm{g}})$, where $\ket{\tilde{\psi}_\mathrm{ex}(t,z^*)}$ is evolved according to the non-linear equation Eq.~(\ref{NonLinHOPSEq}).
To perform expectation values one uses the normalized states \begin{equation}
    \ket{\tilde{V}_\eta(t,z^*)}=\frac{\ket{\tilde{v}_\eta(t,z^*)}}{
     ||\tilde{v}_\eta(t,z^*)||}
\end{equation}
with
\begin{equation}
\begin{split}
   ||\tilde{v}_\eta(t,z^*)||^2
   &\equiv
   \braket{\tilde{v}_\eta(t,z^*)}{\tilde{v}_\eta(t,z^*)}
   \\
   &=
    \frac{1}{2}\left[\braket{\tilde{\psi}_\mathrm{ex}(t,z^*)}{\tilde{\psi}_\mathrm{ex}(t,z^*)} + 1\right].
    \end{split}.
\end{equation}
Note, that the last expression is independent of $\eta$, which means that for all four initial state one has the same normalization factor.
Therefore, repeating the steps that leads form  Eq.~(\ref{eq:C(t)_diag}) to Eq.~(\ref{eq:C(t)_final}) we now arrive at
\begin{equation}
\label{eq:C(t)_final_normalized}
C(t)=  \mu_\mathrm{tot}^2 \mathcal{M}\Big[\frac{\braket{\psi_\mathrm{ex}}{\tilde{\psi}_\mathrm{ex}(t,z^*)}}{\frac{1}{2}\left(||\tilde{\psi}_\mathrm{ex}(t,z^*)||^2 + 1\right)}\Big]e^{i \epsilon_{\mathrm{g}} t}.
\end{equation}
We emphasize, that here $\tilde{\psi}_\mathrm{ex}(t,z^*)$ is propagated with the nonlinear Eq.~(\ref{NonLinHOPSEq}) in the excited state manifold only, but using expectation  values $\langle L_n^\dagger \rangle_t$ calculated with respect to the normalized state $\ket{\tilde{\psi}_\mathrm{ex}(t,z^*)}/\sqrt{({||\tilde{\psi}_\mathrm{ex}(t,z^*)||^2 + 1})}$, which contains a ground state contribution in the normalization.

\section{Numerical calculations\label{sec:calculations}}

In this section, we investigate the numerical performance  of our dyadic equation (Eq.~(\ref{eq:C(t)_final})) using both the linear (Eq.~(\ref{LinHOPSEq})) and non-linear (Eq.~(\ref{NonLinHOPSEq})) HOPS equations and compare their convergence with respect to the number of trajectories. 
In the following, we will calculate the linear absorption spectra for a homodimer ($\epsilon_n=\epsilon$) with electronic coupling $V=1$, where the transition dipoles of both chromophores are parallel. 
We use a simple bath correlation function 
$
\alpha_n(\tau)=\alpha(\tau)=p\,e^{-i\Omega\tau-\gamma|\tau|}
$
where $\gamma=0.25$. Here, and in the following, we choose $\Omega$ as the unit of energy and express $p$ in units of $(\Omega)^2$. The truncation constant for the HOPS, $\mathcal{K}$, defined after Eq.~(\ref{LinHOPSEq}), is provided in the respective plots and is always chosen large enough to be well converged in the hierarchy. 

\begin{figure*}[]
\centering
\includegraphics[width=16cm]{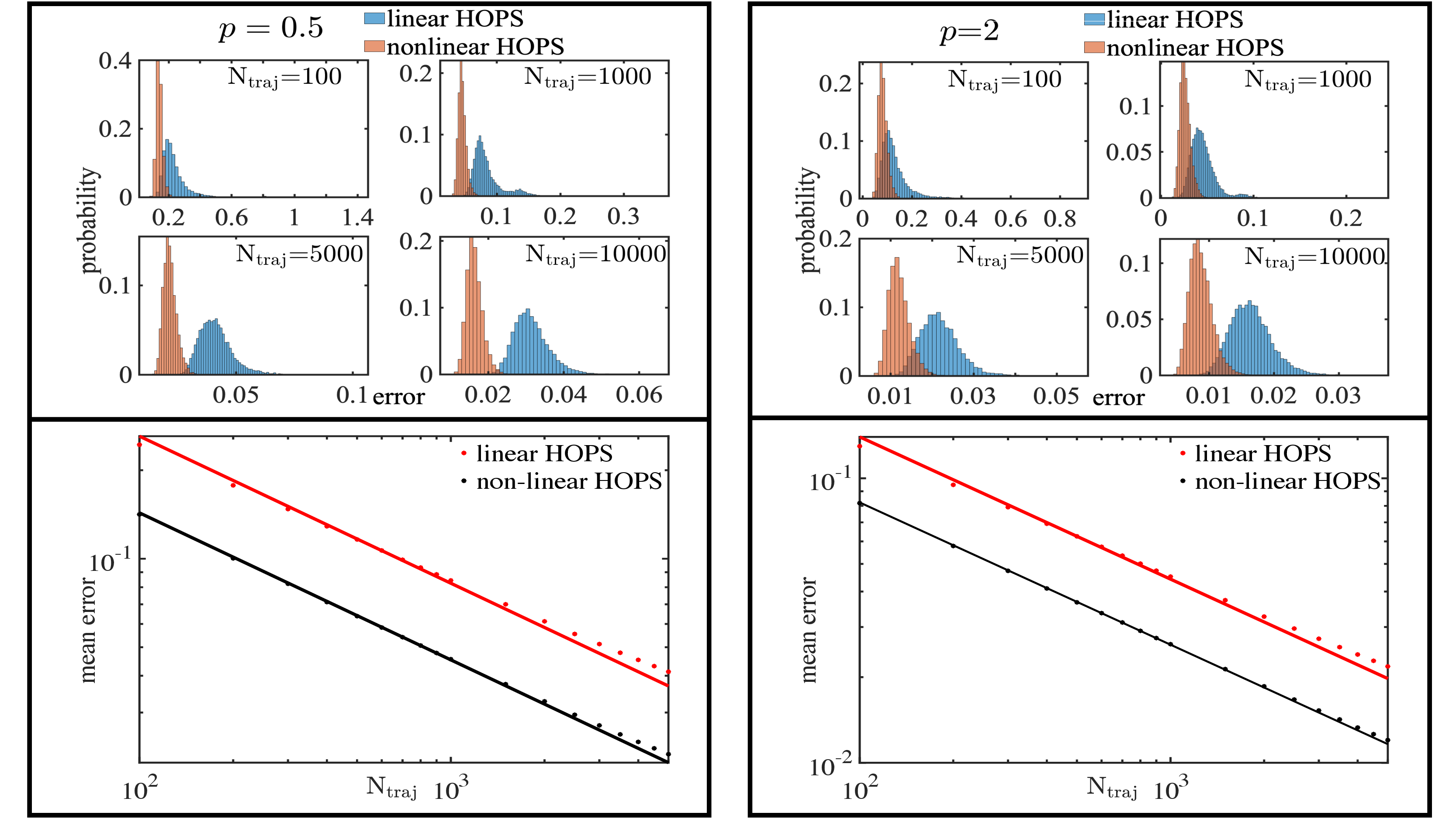}
\caption{Analysis of the error of the stochastic equations compared to the noiseless reference. We consider the same parameters as in Fig.~\ref{fig:Example_spectra_1}. 
Top: The distribution of errors for four different numbers of trajectories, for both the linear and the non-linear HOPS.
Bottom: The mean error of $\mathcal{A}(\omega)$ as a function of $\mathrm{N_{traj}}$.
The straight lines are guides for the eye and indicate a scaling of the mean error as $r_0 \sqrt{ N_\mathrm{traj}}$
}
\label{fig:error_analysis}
\end{figure*}

In Fig.~\ref{fig:Example_spectra_1} we demonstrate that both linear and non-linear HOPS are reasonably converged using a rather small number of trajectories for both $p=0.5$ (left box) and $p=2$ (right box). For both cases the spectrum of the monomer is shown in the top row. In the second row the numerically calculated correlation function of the dimer are shown, for both the linear  and non-linear  HOPS along with the reference spectrum calculated using the noise-free HOPS algorithm. 
The bottom row shows the corresponding absorption spectra.

We quantify the convergence with respect to the number of trajectories using non-parametric error estimation and find the stochastic HOPS calculations converge roughly with $1/\sqrt{N_\mathrm{traj}}$, with the non-linear HOPS showing overall faster convergence. We quantify the difference between the stochastic ($\mathcal{A}(\omega)$) and reference ($\mathcal{A}_{\mathrm{ref}}(\omega)$) spectrum using 
\begin{equation}
\label{eq:diff_measure}
    \mathrm{error}=\frac{1}{\omega_\mathrm{max}-\omega_\mathrm{min}}\int_{\omega_\mathrm{min}}^{\omega_\mathrm{max}} \big|\mathcal{A}(\omega)-\mathcal{A}_{\mathrm{ref}}(\omega)\big|\mathrm{d}\omega
\end{equation}
where the integration extend over the region of $\omega$ in which one is interested in (we take the $\omega$ range shown in Fig.~\ref{fig:Example_spectra_1}).
Note that according to this definition the error is proportional to $\mu_\mathrm{tot}^2$, which in the case shown is equal to 2.
In Fig.~\ref{fig:error_analysis} (upper panels) we compare the distribution errors calculated from average spectra for different ensembles with fixed $N_\mathrm{traj} = 100$, $1000$, $5000$, and  $10000$ (see appendix~\ref{sec:bootstrapping} for details). 
In all cases the non-linear HOPS calculation has error that is approximately half that of the linear HOPS and both have a slightly asymmetric distribution with a full width at half maximum roughly half of their mean. 
In the bottom row of Fig.~\ref{fig:error_analysis} we show the mean values as a function of $N_\mathrm{traj}$ on a log-log scale, where the solid line provides a guide for the expected scaling of a mean error proportional to $1/\sqrt{N_\mathrm{traj}}$.

\begin{figure}[]
\centering
\includegraphics[width=8cm]{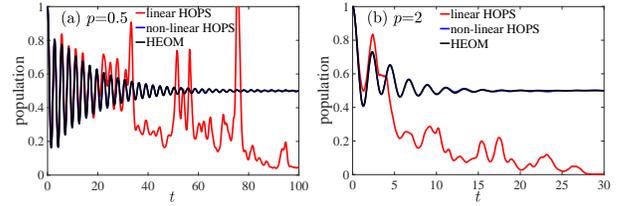}
\caption{Population dynamics of site 1 for $p=0.5$ (a) and $p=2$ (b). The other parameters of the model are the same as in Fig.~\ref{fig:Example_spectra_1} and Fig.~\ref{fig:error_analysis}. 
The number of trajectories used in both linear and non-linear HOPS is 10000. 
One sees clearly that for population dynamics the linear HOPS does not converge for the parameters considered.
The truncation depth of the hierarchy for HOPS and HEOM are $\mathcal{K}=6$ and $7$ for (a), $\mathcal{K}=16$ and $18$ for (b), respectively.
}
\label{fig:Pop_conv}
\end{figure}
These results demonstrate that both the linear and non-linear HOPS can produce accurate correlation function $C(t)$ and absorption spectra $\mathcal{A}(\omega)$. We note that for the strong coupling regime presented here, the linear HOPS converges very slowly for population transfer. 
Fig.~\ref{fig:Pop_conv} compares the population dynamics of site 1 calculated by linear (red line) and non-linear (blue line) HOPS for the two values $p=0.5$ and $p=2$ with the exact HEOM calculations (black line). 
While non-linear HOPS with 10000 trajectories produces very accurate population dynamics, the linear HOPS starts to deviate from the exact results at short time and completely fails to describe the population dynamics at longer time. 
This inability of linear HOPS to describe the strong coupling regime is well known for population dyanmics. 
That linear HOPS can reproduce absorption spectra is a consequence of the cancellation of noise that can be clearly understood in the context of the pure state decomposition.

\section{Conclusions\label{sec:conclusions}}

We derived {\it stochastic} propagation schemes for the calculation of absorption spectra within the NMQSD approach, using both the linear and the non-linear NMQSD equation.
Beside a scheme that relies on formally propagating pure states in the full Hilbert-space, we derived a scheme, where individual propagations in the electroinc ground and electronic excited state are used to obtain the desired spectrum.
This scheme directly resembles the diagrammatic perturbation theory where dyadic matrices are propagated.
While for the linear NMQSD equation the two propagations are completely independent, for the nonlinear NMQSD they become coupled via a common normalization factor that contains the norm of the two wavefunctions in the different Hilbert-spaces. 
Besides its favorable convergence properties with respect to the number of trajectories, this non-linear normalized version will allow a direct implementation of adaptive algorithms.
We investigated the convergence with respect to trajectories $N_\mathrm{traj}$in detail and found that the error decreases as $1/\sqrt{N_\mathrm{traj}}$ with the linear version needing roughly twice the number of trajectories as the non-linear one to achieve the same accuracy.
This behaviour of linear equation is remarkable, since we worked in a parameter regime where population dynamics does not converge for the linear NMQSD  
and our decomposition into pure states does not lead to a   cancellation as was found in Ref.~\onlinecite{HOPSEDRichard}.

We have formulated the absorption spectrum in terms of propagating a single initial state in the electronic excited state.
This state contains all the information of the dipoles (magnitude and orientation) of the individual molecules as well as the local fields at these molecules (for ease of notation we used the same electric field for all molecules).
Since arbitrary distributions of the electromagnetic field can be treated, the formalism can be directly applied to the calculation of circular dichroism or near-field spectra.
We note that similar to the treatment of Ref.~\onlinecite{FTODTFD}, one can also construct the spectrum from an summation of initially localized states, with the correct wheighting factors. 
Such a scheme will be beneficial for an adaptive treatment.

We would like to remark, that our derivations relies on the assumption that decay processes from excited to ground state are negligible (as we have already used in the starting Hamiltonian). 
This is the case in many relevant situations, where these processes are on the nanosecond timescale, compared to a few femtoseconds that are needed to find well resolved spectra.
While such coupling between ground and excited electronic states can be directly treated using the decomposition into pure states and propagating via NMQSD in the full electronic Hilbert space, presumably the dyadic equation will need major modifications. 

We demonstrated the applicability of our schemes explicitly using the HOPS formulation of NMQSD. 
It should be noted that our equation are formulated for arbitrary temperatures.
Within HOPS there are different ways how to incooperate temperature either in the hierarchy \cite{FTODTFD} or in additional stochastic processes.\cite{HOPS2} 
Because our schemes propagate single trajectories subject to noise, it is straightforward to account for the effect of the static disorder induced by the inhomogeneity of the environment without much additional computational cost.
Since the HOPS provides efficient treatment of the environmental degrees of freedom and furthermore one propagates vectors instead of matrices the propagation schemes offer promising techniques to simulate absorption spectra of large molecular aggregates with complicated structured environments.
Our dyadic HOPS equation is also an important step towards application of HOPS for non-linear spectroscopy.

\begin{acknowledgements}
LPC  acknowledges support from the Max-Planck Gesellschaft via the MPI-PKS visitors program. 
 AE acknowledges support from the DFG via a Heisenberg fellowship (Grant No EI 872/5-1).
 DIGB acknowledges support from Robert A. Welch Foundation (Grant N-2026-20200401). 
\end{acknowledgements}

\section*{Data availability}
Further data that support the findings of this study are available from the corresponding author upon reasonable request.

\appendix

\section{Linear absortion using HOPS without noise\label{sec:NoisefreeHOPS}}

It is demonstrated in Ref.~\onlinecite{FTODTFD} that for the calculation of absorption spectra, one can propagate the linear HOPS equation (Eq.~\ref{LinHOPSEq}) without noise, i.e. with  all $z_{t,n}=0$.
The dipole-correlation function is calculated accoring to
\begin{equation}\label{CtSingleTraj}
C(t)=\mu_\mathrm{tot}^2\langle\psi_\mathrm{ex}|\psi^{(\mathbf{0})}_\mathrm{ex}(t)\rangle
\end{equation} 
with
\begin{equation}
\begin{split}
\partial_t|\psi_\mathrm{ex}^{(\mathbf{k})}(t)\rangle= \left(-i\hat{H}_{\mathrm{S}}-\mathbf{k}\cdot\mathbf{w}\right)|\psi_\mathrm{ex}^{(\mathbf{k})}(t)\rangle\\
+\sum_{n}\hat{L}_n\sum_jk_{nj}p_{nj}|\psi_\mathrm{ex}^{(\mathbf{k}-\mathbf{e}_{nj})}(t)\rangle \\
-\sum_{n}\hat{L}_n^{\dagger}\sum_j|\psi_\mathrm{ex}^{(\mathbf{k}+\mathbf{e}_{nj})}(t)\rangle
\end{split}
\end{equation}  
and the initial condition $|\psi_\mathrm{ex}^{(\mathbf{0})}(t\!=\!0)\rangle=\ket{\psi_\mathrm{ex}}$.

\section{Calculations for the Hartmann-Strunz approach}  
\subsection{Writing the initial operator as a sum of pure states\label{sec:diag_operator}}
To obtain equation (\ref{eq:pure_decomp}) we first rewrite 
$\hat\mu_{\mathrm{eff}} \ket{\mathrm{g}}\bra{\mathrm{g}}$ as the sum of two hermitian matrices,
\begin{equation}
\hat\mu_{\mathrm{eff}} \ket{\mathrm{g}}\bra{\mathrm{g}}=\frac{1}{2}\big( \hat{A}+ \hat{B}\big)
\end{equation}
with
\begin{eqnarray}
\hat{A}=&\big\{\hat{\mu}_{\mathrm{eff}}\, , \ket{\mathrm{g}}\bra{\mathrm{g}} \big\}=&\sum_{n=1}^N (\boldsymbol{\mu}_n\cdot\boldsymbol{\varepsilon})\Big( \ket{n}\bra{\mathrm{g}}+\ket{\mathrm{g}}\bra{n}\Big)\\
\hat{B}=&\big[\hat{\mu}_{\mathrm{eff}}\, , \ket{\mathrm{g}}\bra{\mathrm{g}} \big]=&\sum_{n=1}^N (\boldsymbol{\mu}_n\cdot\boldsymbol{\varepsilon})\Big( \ket{n}\bra{\mathrm{g}}-\ket{\mathrm{g}}\bra{n}\Big)
\end{eqnarray}
The eigenvalues and eigenvectors of these two operators can be calculated analytically
\begin{eqnarray}
\hat{A}\ket{v_{\pm 1}}&= &\pm\mu_{\mathrm{tot}}  \ket{v_{\pm 1}} \\
\hat{B}\ket{v_{\pm i}}&= &\pm{i} \mu_{\mathrm{tot}} \ket{v_{\pm i}}
\end{eqnarray}
where the eigenvectores are that defined in Eq.~(\ref{eq:initial_states_diag}) of the main text.

\subsection{Derivation of the dyadic equation\label{sec:calc_dyadic}}
Here we show, that the dyadic equation (\ref{eq:C(t)_dyadic}) is identical to Eq.~(\ref{eq:C(t)_final}).
We start with Eq.~(\ref{eq:C(t)_dyadic})
\begin{equation}
\begin{split}
{\mu}_\mathrm{tot} &\, \mathcal{M} \Big[ \mathrm{Tr}_\mathrm{S}\Big\{\hat{\mu}_\mathrm{eff} \ket{\psi_\mathrm{ex}(t,z^*)}\bra{\mathrm{g}(t)}\Big\}\Big]
\\
&= 
{\mu}_\mathrm{tot} \, \mathcal{M} \Big[ \mathrm{Tr}_\mathrm{S}\Big\{\hat{\mu}_\mathrm{eff} \ket{\psi_\mathrm{ex}(t,z^*)}e^{i \epsilon_{\mathrm{g}} t}\bra{\mathrm{g}}\Big\}\Big]
\\
&= 
{\mu}_\mathrm{tot} \, \mathcal{M} \Big[ \bra{\mathrm{g}}\hat{\mu}_\mathrm{eff} \ket{\psi_\mathrm{ex}(t,z^*)}e^{i \epsilon_{\mathrm{g}} t}\Big]
\\
&= 
{\mu}_\mathrm{tot}^2 \, \mathcal{M} \Big[ \braket{\psi_\mathrm{ex}}{\psi_\mathrm{ex}(t,z^*)}e^{i \epsilon_{\mathrm{g}} t}\Big]
\end{split}
\end{equation}
where  in the first step we used $\bra{\mathrm{g}(t)}=e^{i\epsilon_{\mathrm{g}} t}\bra{\mathrm{g}}$, in the second step we evaluated the trace and in the last step we have used Eq.~(\ref{eq:psi_ex}). 
We see that the last line is identical to Eq.~(\ref{eq:C(t)_final}).

\section{Analysis of the statistic error by the bootstrapping method\label{sec:bootstrapping}}
To get a quantitative description of the convergence properties of linear and non-linear HOPS with respect to the number of trajectories, we conduct a detailed analysis of the statistic error due to a finite number of trajectories by using the bootstrapping technique \cite{BootStrap}. 
To this end, we first calculate $3\times{10}^4$ trajectories. 
For each value of $\mathrm{N_{traj}}$, we construct $10^4$ ensembles by selecting $\mathrm{N_{traj}}$ trajectories randomly from the original $3\times{10}^4$ trajectories (the same trajectory can appear multiple times within each ensemble). 
For each ensemble, we quantify the error by evaluating the average absolute difference of $\mathcal{A}(\omega)$.



\begin{thebibliography}{}
\bibitem{NMQSDfootnote} This equation is equivalent to $
C(t)= \mathcal{M} \Big[ \mathrm{Tr}_\mathrm{S}\{\hat{\mu}_\mathrm{eff}\, (\hat{U}(t,\mathbf{z}^*) \hat\mu_\mathrm{eff} \ket{\mathrm{g}}) \bra{\mathrm{g}}\hat{U}^\dagger(t,\mathbf{z}^*)\} \Big]$
where $\hat{U}(t,\mathbf{z}^*)$ denotes the stochastic propagator in the complete Hilbert space.
\end{thebibliography}
\end{document}